# WHEN GENERATIVE AI MEETS WORKPLACE LEARNING – CREATING A REALISTIC & MOTIVATING LEARNING EXPERIENCE WITH A GENERATIVE PCA

*Completed Research Paper*

Andreas Bucher, University of Zurich, Zurich, Switzerland, bucher@ifi.uzh.ch

Birgit Schenk, University of Public Administration and Finance Ludwigsburg, Ludwigsburg, Germany, birgit.schenk@hs-ludwigsburg.de

Mateusz Dolata, University of Zurich, Zurich, Switzerland, dolata@ifi.uzh.ch

Gerhard Schwabe, University of Zurich, Zurich, Switzerland, schwabe@ifi.uzh.ch

## Abstract

*Workplace learning is used to train employees systematically, e.g., via e-learning or in 1:1 training. However, this is often deemed ineffective and costly. Whereas pure e-learning lacks the possibility of conversational exercise and personal contact, 1:1 training with human instructors involves a high level of personnel and organizational costs. Hence, pedagogical conversational agents (PCAs), based on generative AI, seem to compensate for the disadvantages of both forms. Following Action Design Research, this paper describes an organizational communication training with a Generative PCA (GenPCA). The evaluation shows promising results: the agent was perceived positively among employees and contributed to an improvement in self-determined learning. However, the integration of such agent comes not without limitations. We conclude with suggestions concerning the didactical methods, which are supported by a GenPCA, and possible improvements of such an agent for workplace learning.*

## 1 Introduction

Establishing the workplace as a learning environment and fostering the development of employees' skills and knowledge is crucial for an organization's success and competitive advantage (Caputo et al., 2019; Nagano, 2020). Especially in times of an increased shortage of high-skilled labor and the increasing automation of tasks due to digitalization and AI, organizations need to rethink their approach to workplace learning. Generative AI (GenAI), and in particular pedagogical conversational agents (PCAs) that build on large language models (LLMs), appears as a promising solution to enrich the learning process and to provide didactical methods that were previously reserved for instructor-led training to a larger workforce (Abdelghani et al., 2023; Sharples, 2023). This study explores how a generative AI-based PCA or simply GenPCA can support workplace learning, particularly a communication training of two public administrations in Germany, and how such training is perceived by the employees.

To ensure a high level of citizen service, organizations define communication guidelines, like a mandate to return missed calls within two days and ensure continuous phone line availability (Schenk and Gaeng, 2022). Some organizations even offer 1:1 training sessions for new employees, led by Human Resources (HR), to train employees and raise awareness for good citizen service. However, COVID-19 sent employees to remote work overnight without ensuring service provisioning. Also, previous training programs were halted due to workforce shortage. As a result, service quality dropped significantly.

This was also experienced by the two public administrations of this study. In 2022, the two organizations thus decided to renew their efforts to foster a citizen-centric service culture, necessitating the training of approximately 600 staff members each who interact directly with the public. Nonetheless, it soon became apparent that providing 45-minute training sessions for each of these employees would be infeasible for the HR departments. Additionally, the logistical challenges of scheduling and follow-up were significant. This prompted the exploration of an alternative solution: the development of an e-learning program that could be accessed flexibly at the workplace.

However, questions arose about how to enrich the training with interactive exercises and the feasibility of guiding employees through the program without a human trainer. Previous experiences with PCAs in educational settings (Wambsganss et al., 2020, 2021; Winkler et al., 2019; Winkler and Soellner, 2018), promised to be a potential substitute. Additionally, the wake of GenAI, especially the wide-reaching capabilities of LLMs, promises new ways of interacting with learners in a highly immersive and natural learning environment (Abdelghani et al., 2023; Sharples, 2023). Yet, the application of PCAs in workplace learning, in particular the use of GenPCAs to train employees in their communication skills, remains under-researched. This led to the following research questions:

**RQ1**: *Can similar didactical methods for workplace learning be applied in GenPCA-based training as in training with a human instructor?*

**RQ2**: *What are the effects of GenPCA-based training on motivational factors of workplace learning?*

To answer these research questions, we engaged in a research project with two German public organizations to train employees in their communication skills using a GenPCA, called DIMA. DIMA is able to interact with the learners through a chat on an e-learning website, e.g., to answer questions related to the overall training process and content, and can conduct practical exercises with the learners over telephone and E-Mail, e.g., to simulate a frustrated citizen. Over a period of 12 weeks, 63 employees trained with the prototype and evaluated it. Based on 32 interviews with these employees, we found that a GenPCA-based training can provide a realistic and motivating learning setting, which reduces the need for human instruction and supports the development of the learners' communication skills.

## 2   Related Work

### 2.1   Workplace learning & didactical background

Knowledge development and transfer is one of the most important tasks in an organization. The shortage of highly skilled employees and the need to upskill and reskill them due to increasing digitalization and automation of tasks requires many organizations to rethink workplace learning (Morandini et al., 2023; Noe et al., 2010). To create effective workplace learning programs, it is important to consider the high-level and structural learning processes and to fulfill the basic principles of adult learning (Illeris, 2003). Adult learning is commonly referred to as *andragogy*, which defines learning as a self-directed inquiry (Knowles et al., 2020). Therein, adult learners need to control the nature, timing, and direction of the learning process and to understand the meaning and importance of what is being learned (Illeris, 2003; Rogers, 1969; Russell, 2006). As a lack of motivation and engagement is a commonly encountered issue in workplace learning, considering motivational aspects in the design of workplace learning programs is essential. According to Deci and Ryan's needs-based *self-determination theory* (SDT) of motivation (Ryan and Deci, 2000), people's intrinsic motivation for engaging in learning activities is determined by the satisfaction of their three basic psychological needs for autonomy, competence, and relatedness.

Carefully selecting appropriate didactical approaches influences and supports motivation. For instance, employees' perception of competence can be improved through *adaptive learning*, which adapts the learning process to peoples' prior experiences, their skill level or job relevance (Karabenick, 2014; Plass and Pawar, 2020). By *integrating theory and practice* and by providing *experiential learning*, which is challenging but not overwhelming, workplace learning programs can stimulate the learning experience and put the learners into a zone of proximal development (ZPD) (Kolb and Kolb, 2009; Noe et al., 2010;

Wang, 2018). Similarly, offering corrective *feedback and assessment*, which can be challenging for untrained instructors (Ingram et al., 2013), enhances the learners' perception of competence (Ryan and Deci, 2000). Additionally, enabling *self-directed learning* by offering choices throughout the learning process and putting the employees in charge of the time and location of the learning, empowers employees and evokes feelings of autonomy (Lemmetty and Collin, 2020; Taylor, 2006). Lastly, *social learning*, which connects learners and establishes an environment and culture for learning, can satisfy the learners' needs for relatedness and social presence (Minnaert et al., 2011; Oh and Song, 2021).

To promote a collective understanding of the training's value, encouraging autonomy and self-driven learning, requires careful planning. A variety of strategies can be used to improve the transfer of knowledge among employees and to foster a common comprehension of educational materials. These strategies range from traditional in-person training led by instructors to e-learning courses (Martins et al., 2019; Shaw et al., 2009). Researchers and practitioners have devoted a lot of attention to the design of instructor-led training, which is the predominant framework in organizational training (Noe et al., 2010). Many approaches exist to guide the instructional design of workplace learning: for the acquisition of theoretical knowledge, various expository (e.g., lectures or presentations) and inquisitory methods (e.g., tutorials) are available (Womack, 1989). To internalize new skills and knowledge, instructors can apply practical (e.g., drills or simulations) and interactive exercises (e.g., discussions) and provide reflections and feedback (Gilmore and Fritsch, 2001; Molenda et al., 2006). An important aspect in instructor-led training is the instructor's ability to correctly assess the learners' emotions (Siu and Wong, 2016; Urhahne and Zhu, 2015). However, instructor-led trainings are often criticized due to their instructor-centricity leaving the learners often as passive recipients of information, which undermines learners' motivation (Noe et al., 2010). In contrast, e-learning is perceived as more learner-centric; although, it commonly fails to engage and motivate learners (Noe et al., 2010). We believe that applying GenPCA-based training in professional learning programs, which can potentially replace human instructors and enhance e-learning, is a promising direction to improve workplace learning.

## 2.2 Pedagogical conversational agents

PCAs are the intersection of different research streams and combine aspects of pedagogical agents and intelligent tutoring systems with conversational agents (CAs). PCAs emphasize the conversational character of virtual agents to guide the learning process (Johnson and Lester, 2018). A common issue in education is related to a lack of learner engagement leading to boredom and demotivation. Hence, research on PCAs focused on systems that provide personalized and adaptive learning paths and teaching to account for learners' distinct backgrounds and characteristics (Kabudi et al., 2021). PCAs are effective in a variety of educational contexts, e.g., to support the development of better argumentation and reasoning skills (Wambsganss et al., 2020, 2021), to enhance students' problem-solving skills (Winkler et al., 2019), or to teach programming skills with a scaffolding-based PCA (Winkler et al., 2020). Similarly, the educational benefits of PCAs, especially due to their interactive and personalized character, have been proven effective in language learning (Lee and Lim, 2023), in preparing high-school students for tests (Waldner et al., 2022), or in training medical students (Al Kahf et al., 2023). Although many studies support the notion of the educational benefits of PCAs, only a few studies have investigated the use of PCAs outside of the academic and educational context. Nonetheless, PCAs are also beneficial in training and upskilling adults, e.g., within the scope of organizational training programs (Bendel, 2003; Curtis & Thomas, 2008; Meyer von Wolff et al., 2019). For instance, Bucher et al. (2023) found that PCAs are effective for training telephone service in organizations although they evoked feelings of surveillance. Other studies (Kocielnik et al., 2018; Wolfbauer et al., 2023) explored how PCAs can be used by employees to reflect on their actions and behavior at the workplace to foster learning. Despite their potential for workplace learning, however, research in this area is still very nascent.

When developing PCAs, the design questions are similar to those that arise during the development of general-purpose CAs. Therefore, existing design guidelines and taxonomies (Feine et al., 2019; Zierau et al., 2020) for the development of CAs can offer guidance in design choices; however, the design of PCAs must also specifically account for the characteristics of the learning context. To guide design choices, Wellnhammer et al. (2020) and Weber et al. (2021) provide an overview of different design elements that need to be considered. Additionally, the *Learning with Pedagogical Agents* (LPAM) model by Dolata et al. (2023) illustrates design considerations for the PCA and their relation to the learning environment based on the activity theory. Such design considerations span from the virtual representation or embodiment (Kim et al., 2018) to the integration of a PCA in e-learning or ITS environments (Cook et al., 2017; Keller and Brucker-Kley, 2020; Kulik and Fletcher, 2016) or their specific role (Kim and Baylor, 2016). The effects of these design considerations are wide-reaching and should be considered carefully. Some studies suggest that PCAs should offer explanatory and spoken feedback (Heidig and Clarebout, 2011), or should augment human-instructors in the learning process (Sjöström and Dahlin, 2020). Therein, a PCA can take over various roles, such as an expert, teacher, tutor, or peer and sparring partner (Chen et al., 2020; Kim and Baylor, 2016; Tegos and Demetriadis, 2017)(Chen et al., 2020; Kim and Baylor, 2016; Tegos and Demetriadis, 2017). Combining and switching between different roles, such as acting as a tutor for explanations and as a peer for training, can further enhance the learning outcome (Chen et al., 2020). Additionally, research started exploring video-based detection to assess the learners' state of mind to adapt the learning process (Mukhopadhyay et al., 2020). By integrating human-like features, PCAs establish a social presence and personal bond with their users (Khosrawi-Rad et al., 2023; Lester et al., 1997) or create a particular power distance (Schlimbach and Zhu, 2023). However, as workplace learning differs quite greatly from the educational context, e.g., in terms of learners' characteristics and the organizational setting, it remains unclear how previous findings are transferable to workplace learning.

Often, the interaction with PCAs is not perfect. Similar to common conversational breakdowns with CAs, PCAs can often not provide adequate support and do not immerse learners in engaging conversations (Janssen et al., 2021; Khosrawi-Rad et al., 2023; Tolzin et al., 2023). However, recent advances in large language models, such as the introduction of OpenAI's GPT-4 (OpenAI, 2023), Google's PaLM 2 (Anil et al., 2023), or Meta's Llama 2 (Touvron et al., 2023), offer new ways for interacting with a PCA. Such models have great consequences for peoples' lives as they can augment and assist humans or even replace human labor (Dwivedi et al., 2023). Based on LLMs, the capabilities and scope of PCAs increase likewise. GenPCAs, based on LLMs, can take over new roles in the learning process, such as a Socratic opponent or a collaboration coach (Sharples, 2023). Such PCAs inhibit advanced information processing capabilities and enable new forms of natural communication. For example, a GPT-3-based PCA can generate linguistic and semantic cues to help learners' divergent questioning skills and foster curiosity (Abdelghani et al., 2023). GenPCAs offer the potential for new ways of engaging learners, however, research on such agents is still nascent. For instance, it remains open how advanced dialog capabilities of GenPCAs affect the application of didactical methods and the perception of GenPCAs. Also, integrating such agents beyond the educational context and into adult learning is unresearched.

## 3 Methodology

The intention behind this study is to identify and address real-life problems. For this, Action Design Research (ADR) (Sein et al., 2011) provides a suitable framework. ADR combines design activities (= building and evaluating artifacts) and action research (= intervening and evaluating interventions). Its steps include problem formulation, conducting the artifact-based intervention, evaluating it, reflection and learning, and, finally, formalization of the learning. This study focuses on problem formulation through fieldwork and conceptualization, and on specifying a potential intervention through reciprocal and user-engaged design. We describe the findings of our study following Lee et al. (2011).

Whereas real problems and solutions can be observed and tested in the *instance domain*, the *abstract domain* allows for theorizing and scientific discourse. Our generalizable 'abstract problem' (Lee et al., 2011) is managing the shortage of training personnel in public administration to provide adequate workplace learning. It is instantiated by the lack of trainers for personalized communication training. The example of communication training was selected because it is essential to improve service quality, which is a relevant complex problem for many public administrations. All municipalities need a high service quality to present themselves to citizens as an entity that is trustworthy and reliable. The research was initiated by one city that asked for help in enhancing their service quality. Furthermore, when conducting an informal pre-study, we received signs of interest from another cooperating municipality. These two municipalities are interested in digitalizing their workplace learning to meet their problem of 1:1 training. Both are pioneers through specific research projects in the field of e-government. To derive the instance solution, which involved the development of a GenPCA called DIMA (= Digital Intrinsic Motivation Agent), we first defined the didactical and methodological framework for a training program with such agent. This framework included aspects of adult learning and SDT as well as instructional methods for training employees in their communication skills.

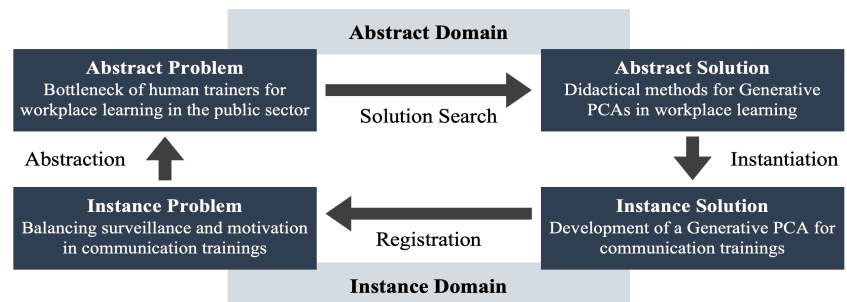

*Figure 1.*     Methodological approach following Lee et al. (2011)

## 3.1     Design of the workplace training

Since this study is situated in a larger research project, the training program draws on insights from prior trainings and discussions with experts. It was tailored to the requirements of adult learning and SDT. Overall, the training comprised nine modules to teach various aspects of communication. Starting with the introduction of the corporate identity, the learners were introduced to handling telephone calls with citizens and colleagues and written communication (e.g., E-Mails). To encourage self-determined learning, participants could choose: a) where and when they learn, b) the order of the units, and c) the number and frequency of exercises. They can also decide when to contact DIMA during the training. The training's significance was outlined on the landing page of an e-learning platform, emphasizing both its general aim and specific relevance to administrative work. Each unit began with an explanation of its importance and objectives, underscoring the practical application in the workplace.

The training was structured into nine concise segments, each fitting into the work schedule with the flexibility to take breaks as needed. Developed in collaboration with representatives of the administrations, each unit spans 20-30min. To facilitate the application of learned skills to real tasks and enhance motivation, the units featured various exercises based on real-life scenarios from public administrations. The complexity level varied across units to cover different skills, such as general conversation or handling conflict situations. In prior 1:1 training with human instructors, they took over the role of a tutor and provided instructions, general feedback, and guidance throughout the training. Engaging with employees in practical exercises typically involved a human sparring partner as the instructor assessed the learning situation for subsequent feedback. Both roles, tutor and sparring partner, represented the basis for our GenPCA.

## 3.2     Design of DIMA based on principles of self-determined learning

DIMA's design, rooted in principles of self-determined learning, aimed to meet the learners' needs for autonomy, competence, and relatedness. Building on our research (Bucher et al., 2023), we found PCAs effective for upskilling but noted that ignoring employees' needs for autonomy can feel like surveillance, hindering learning. To avoid the perception of surveillance, we took special considerations of the learners' *autonomy* by placing DIMA into a formal learning setting supported by an e-learning platform. Also, unbundling the learning process from human involvement provides more flexibility as learners can independently conduct exercises and do not rely on a human trainer for explanations. This enables the learners to select the time and location, and also the sequence of the learning process. By adapting the exercises with DIMA in the role of a sparring partner to real-world scenarios and by providing individualized feedback on the performance of learners during the exercises, DIMA should support the employees' need for *competence*. Lastly, the selection of specific social cues and interaction modalities, such as a human-like voice, name, or gender, and the ability to act intelligently to employees' input should create a sense of presence with DIMA and support the need for *relatedness*.

DIMA fulfills two different roles during the training: the role of a tutor and the role of a sparring partner. Both roles encompass distinct design elements, social cues and tasks. For instance, in the role of the tutor, DIMA's tasks were to provide guidance throughout the training by answering questions related to the learning process and material, and to offer feedback on exercises that were conducted with the sparring partner. In this role, employees interacted informally with DIMA via chat, E-Mail or telephone. Therein, learners could choose the gender of the tutor, which also determined its voice. When engaging as a sparring partner, DIMA simulated a citizen or a colleague. For instance, the learners had to solve a request from a citizen or explain specific communication rules to a colleague. The interaction with the sparring partner was rather formal when acting as a citizen and informal when acting as a colleague.

DIMA was engineered with a modular architecture, comprising three separate subsystems: the *Q&A agent module*, the *Telephone agent module*, and the *E-Mail agent module*. The Q&A agent module served as main interface for learners, allowing employees to interact with DIMA through a chat feature on the e-learning platform. The e-learning platform was hosted by the provider Howspace and was adapted to our training. Here, DIMA offered answers to queries about the training material, scheduled training sessions, and initiated exercises, which are then carried out by the Telephone or E-Mail agent. The Q&A agent module utilized OpenAI's GPT-3.5 API and a series of parallel and sequential API calls, optimizing response times and enabling a seamless flow of structured yet natural dialogue. Further system prompts confined conversational topics to the training context and reduced hallucinations. Initiating a telephone-based exercise via the Q&A agent triggered the Telephone agent module to set up a call using Twilio. This call was transcribed using IBM's speech-to-text (STT) technology and analyzed with the help of OpenAI's GPT-3.5 and GPT-4 models. Exercises included various prompts that were dynamically constructed throughout an exercise to account for the two distinct roles of DIMA. In the role of a tutor, DIMA was continuously prompted as a friendly and supporting learning assistant. In contrast, as a sparring partner, DIMA was prompted according to the specific exercise (e.g., as an angry citizen). DIMA's responses were then processed by Twilio's automatic text-to-speech (TTS) generation.

### 3.3   Evaluation & analysis

Overall, 63 employees participated in our training program and interacted with DIMA over the course of 12 weeks. Of these 63 employees, 63.5% associated themselves as female, the majority were older than 38 years old and had more than 15 years of work experience in the public sector at the time of the study. Despite working in different departments such as citizen service, building authority or tax office, all had regular contact with citizens or other external entities. After 12 weeks, 46 employees (73%) completed the training. Of those employees who completed the training program, we interviewed 32. Those who did not finish the training or did not participate in the interviews cited reasons such as contract termination, parental or long-term sick leave, or insufficient time due to workforce shortages. At the time of the interviews, 59% of interviewed employees identified as female and were on average 33.5 years old. As no additional themes appeared after the 23$^{rd}$ interview, we concluded to have reached data saturation (Saunders et al., 2018) based on our analysis. The interviews were conducted and recorded using MS Teams and lasted on average 29min. They covered four larger topics: 1) the overall perception of the training, 2) the learners' experiences when interacting with DIMA using the chat, and their impression of the exercises with DIMA, which were conducted over 3) telephone and 4) E-Mail.

We analyzed the interviews using a bottom-up approach (Saldaña, 2013) and divided them into smaller, thematically related segments. Initially, we focused on four areas: learners' perceptions of DIMA in its role as 1) tutor and 2) sparring partner, 3) how DIMA influenced the training, and 4) what challenges arose during training with DIMA. Through several rounds of coding and further analyses by two re-searchers, the final set of 20 codes was derived, which include, e.g., the perceived formality of communication or hedonic qualities of DIMA. The categorization and analysis of the coded segments identified patterns and recurring themes, which provided a deeper understanding of employees' perceptions, the impact of the training, and any limitations associated with it. To ensure reliability and validity, we conducted member checks, peer reviews and examinations, and audit trails (Merriam and Grenier, 2019).

## 4 Results

To better understand the underlying mechanics and effects of GenPCA-based training, the employees' specific perceptions of the GenPCA, including its social cues, roles, and interaction modalities, and their general perception of a GenPCA-based training are presented in the following. As our GenPCA DIMA took on two different roles, that of a tutor and that of a sparring partner, the findings will be separately outlined. Generally, valuable feedback on DIMA's roles emerged from the interviews, alongside general perceptions of the training's impact on communication skills and knowledge. The following summarizes key observations from the interviews, supported by participant quotes (labeled *P1* to *P32*).

### 4.1 Perception of a GenPCA in the role of a tutor

As a tutor, DIMA fulfilled multiple tasks, including guidance throughout the learning process, answering questions concerning the learning material, and providing feedback on the employees' performances during the practical exercises. Overall, the learners perceived DIMA and its support during the training positively. Therein, the employees perceived DIMA as a learning partner and not like a teacher or trainer. For instance, P6 mentioned *"..., it was more like learning together. So, it wasn't someone telling me how or what I should do, what I had to do, but rather someone saying, okay, you could do this, and then giving me feedback. So, I would rather say learning partner."* It is important to highlight that the perception of the learners is not solely determined by the underlying LLMs for response generation, but by the interplay between different elements. For instance, when learners interacted with DIMA over the telephone their perception was likewise influenced by STT translation and TTS generation. Hence, multiple elements, including the agent's ability to converse in a human-like form including its ability to flexibly process and generate natural language, the informality of the speech-situation, and anthropomorphic and social cues, established a motivating and trustful learning environment.

Overall, all employees ascribed **human characteristics** to DIMA. These characteristics range from friendly and understanding to attributes like honest and intelligent. Although the employees knew that DIMA was artificial, they did not perceive DIMA as a computer program. For instance, P6 described that when interacting with DIMA *"I didn't have the feeling that I was talking to a computer, so of course you could hear it vocally and everything, but I didn't have the feeling that I was really talking or writing to a computer all the time, it was human."*. Other employees were more reserved and called DIMA an *"electronic person" (P8)* or *"like a robot" (P14)*. Also, some missed that DIMA could not understand implicit dynamics and interpersonal matters. For instance, P8 described that *"it would be nice, of course, if it was able to take the interpersonal aspects into account a little"*. Besides this, many employees still mentioned a form of social bond that formed with DIMA, such as P10 who reported that *"After a while I had that certain bond, because it was familiar."*. However, some employees also focused on the learning task at hand and perceived DIMA simply as a learning tool or explained that DIMA would affect their emotional level less than a human trainer. For example, P2 noted that *"...the fact that you don't actually have to deal with them directly opposite you, for example, means that the emotional level is less affected."*. Nonetheless, the employees valued the human-like form of DIMA as it was engaging and provided a learning environment in which they felt *"well looked after and well guided"* (P4).

As the primary interaction modality involved natural language, DIMA's **dialog and speech capabilities** had a profound influence on employees' perception of the agent. In general, the learners perceived the rather informal language of the agent as pleasant. For instance, P1 explained that she *"actually found [informal] pleasant, it was so direct that I didn't have to pay too much attention to any polite phrases and so on ...but could simply ask a question directly"*. Apparently, using informal language, including calling the employees by their first name, contributed to a lower power distance with DIMA and less reluctance of interacting with the agent. Additionally, using voice-based interactions with DIMA during the exercises, e.g., when receiving feedback on the tasks, made employees feel more comfortable when interacting with the agent. However, the dialog and speech capabilities did not come without limitations. Sometimes, response times were too long, or transcripts of the conversations were inaccurate resulting in a propagation of errors throughout DIMA's natural language understanding and response generation. In rare cases, this resulted in misunderstandings and hallucinations including the provision of inaccurate information about organizational practices. Also, despite relying on state-of-the-art speech generation services, the employees still perceived the voice of DIMA as too robotic. Therein, P15 concluded that the voice *"sounded a bit like a computer"* whereas P*9* mentioned that *"it was actually pleasant, very calm. It's still a bit mechanical, so as I said, it lacks a bit of tone or dialect or something like that, but otherwise it was actually pleasant."*. Followingly, some employees wished for better language understanding and less mechanical voices that entail dialect or more intonation.

Lastly, most employees valued the capability of the agent to **dynamically adapt its responses** to their requests. Therein, they ascribed some form of intelligence to the agent. Additionally, many employees perceived the adaptability of the agent as a benefit for effective communication and knowledge transfer. For example, P6 compared DIMA's responses with common chatbots or telephone service agents: *"If you don't get these typical answers that you get when you ask a question in a chat, I don't know, on the phone service or something, then you usually get these typical answers that don't help you that much. So that was different, it came across as more human to me."*. Similarly, P16 stated *"I really liked the fact that he [DIMA] gave a personalized answer. The fact that it wasn't just some pre-prepared sentences, yes, I thought that was quite accurate and helpful."*. Together with other social cues, the adaptability of the responses contributed to a feeling of closeness and naturalness.

### 4.2   Perception of a GenPCA in the role of a sparring partner

When conducting exercises with DIMA, the GenPCA acted as a sparring partner, e.g., by taking the perspective of a simulated citizen or a colleague. To create a realistic impression, we chose specific **real-world scenarios** and tailored the communication to the conversation with the learners. As in one scenario, the learners had to react on the telephone to a frustrated citizen and solve a particular situation (e.g., to calm down a citizen, who was angry about cars that are parked on the sidewalks). Herein, DIMA was prompted to act as a certain character to better convey its role of a sparring partner. Concerning the scenarios, P7 said that *"you could really empathize, and the examples were well done. You have to engage with it [DIMA] and of course ignore the fact that it's not real. If you get into it, then it's very realistically done."*. Similarly, P3 described the scenarios that *"it could have been like that in real life, and I didn't think it was fake and you can really put yourself in the shoes of the person. And that's why I thought it was authentic"*. Nonetheless, some employees described their initial scepsis when interacting with the simulated citizen. They initially believed that the scenarios were too unrealistic, but realized later that the scenarios were actually inspired by real-life cases. Therein, P13 described that *"I found some of the examples quite curious, although I have since learned that these are actually cases where people call the town hall"*. Apparently, the exercises with DIMA did not only train the employees' communication skills but enhanced the awareness for specific needs of the citizens.

Besides offering real-life scenarios, the **adaptivity of the sparring partner** to the current speech situation and its particular behavior contributed to a perception of realism during the training. Building on LLMs, DIMA was not only able to answer specific questions or provide individualized feedback as a tutor, but it could also flexibly adapt the communication of the sparring partner to the flow of conversation. This was observable concerning the communication with the sparring partner over the telephone, but also via E-Mail. Even as the employees were aware that they interacted with DIMA, they described the exercises with the simulated citizen as close to real interactions. E.g., P11 reported about the telephone exercises that *"you could tell it wasn't a real citizen, but it came very close, I have to say"*. When conducting the exercises via telephone, the employees valued the fact that the sparring partner, acting as a citizen, was not easily put off and satisfied with the employees' answers. Herein, P5 mentioned that "*I also thought it was good that they didn't let it go*". Overall, the personalization of the exercises and the adaptivity to the speech situation created a realistic learning context and engaged the employees.

Also, the sparring partner was designed to **express emotions** to create a more realistic experience. To express emotions, like frustration or anger, DIMA was prompted to leverage the LLMs and select its choice of words. Concerning the ability to express emotions, P1 mentioned that *"the way the emotions were conveyed, for example, or the anger or whatever the dissatisfaction or whatever, I felt it was very realistic"*. However, some employees also believed that the interactions with the sparring partner could even be more realistic. When comparing the exercises to real-life interactions with citizens, P12 compared that "*someone who has a similar problem would also speak like that but would be more heated than Dima*". DIMA, however, was not able to change its pitch or speech rate.

### 4.3 General perceptions regarding GenPCA-based training

During the training, the learners had multiple touchpoints with DIMA. Overall, the design of DIMA was intended to support various didactical aspects of workplace learning, including self-directed learning and experiential learning. Through different practical exercises, which were designed for higher learner engagement, the employees should experience an integration of theoretical and practical aspects of learning. As the learners described in the interviews, GenPCA-based training positively impacted different aspects of self-determined learning, enabled an engaging learning environment, and ultimately increased learners' understanding of the learning material and changed their communication behavior.

The objective of the training was to improve the organization's service quality by training in various aspects of communication. The learners reported that training with DIMA, both in the roles of a tutor and sparring partner, positively affected the learning outcome. Generally, the majority of learners rated the training positively and described an **overall improvement** in their communication skills as learning result. For instance, P6 described the training as *"very good in itself. As I said, I already knew a lot of things, but I also learned new things, especially how to deal with clients when things are unclear, which happens very often here, and how to clarify things"*. Furthermore, the learners mentioned that they were able to transfer the learned things into their daily work as noted by P16: *"I was able to transfer a lot of what I learned into my day-to-day work."*. Apparently, the training provided an effective setting for the learners to improve their communication skills and knowledge and to transfer it into their daily work.

A deeper analysis revealed that the GenPCA-based training increased **the level of understanding** for communicative aspects and improved or confirmed employees' **confidence in their communication skills**. During the training, DIMA provided answers to content-related questions and guided the learners through the learning process. Additionally, DIMA gave corrective feedback after conducting an exercise. Providing feedback supported self-reflection and self-awareness, and, thus, individual improvement. On the one side, "*the feedback reminds you again of the important points*" as P2 mentioned. On the other hand, the feedback encouraged self-reflection and provided practical tips. For instance, P9 described *"when I received feedback, [and] that was something I took notes on, there were things I realized, yes, maybe I've been doing this in the same way for years. And it's quite good that he [DIMA] gave me one or two tips to think about it a bit more, and that's just to consolidate it in my head."*. Furthermore, practicing with DIMA also had the effect that employees started feeling more confident in their communication skills as DIMA offered a safe training ground. As described by P16 *"it was really quite good to practice with a virtual person, so to speak. It gave you a bit more confidence and helped."*. Similarly, employees mentioned that engaging with DIMA in E-Mail exercises confirmed their skills. Herein, P18 stated that *"I already felt very confident, but it was nice to get confirmation that I was doing the right thing from Dima's point of view"*. Obviously, the interactions with DIMA not only contributed to a better understanding but also motivated employees to engage in the training by bolstering their confidence in their skills through practice and confirmation.

Providing effective training requires an **engaging and motivating learning environment**. In this regard, the employees valued DIMA's ability to offer practical exercises without the need for human involvement, which fostered their autonomy. For example, P18 said *"I think it's more practical […] if it's not a real person, because Dima is waiting for me. If I say that I don't have time now because something* *happens, [for example] someone comes in, then Dima doesn't care. But a person, well, that wouldn't be possible for a person"*. Additionally, P16 questioned the need for human involvement: *"I really found it quite pleasant and that you can do it at any time. And in terms of the structure, I didn't have the feeling that a human person was needed, so everything was covered"*. Consequently, DIMA's integration allowed for greater flexibility in learning time and location as no human instructor or sparring partner was needed. However, some employees desired more proactive engagement from DIMA to maintain focus on learning despite their need for autonomy. Also, training with DIMA was enjoyed for its immersive and enjoyable atmosphere. The training was perceived as *"cool, a nice change."* (P12) and made the employees curious as P18 stated: *"I really liked the E-Mail training, I found it the most exciting because I was just so curious about what Dima would say specifically about this E-Mail I sent. I really liked that, I was super curious."*. Using a GenPCA-based training apparently provides for an exciting learning experience and positively influences peoples' motivational drivers, like their perception of autonomy.

## 5    Discussion

The findings indicate that a GenPCA-based training effectively enhances employees' communication skills, functioning comparably to human instructors. Such training offers flexibility and accessibility, fostering engaging learning environments. Even though our study focused on workplace learning in public administrations, there are notable similarities with other types of organizations in terms of associated challenges of employee motivation, and the cost and time involved in implementing workplace learning programs. Hence, we believe that our findings can also be transferred to workplace learning in other professional contexts. To answer the research questions, we derive five didactical methods for GenPCA-based training and discuss the impact of such training on employees' motivation. Lastly, the limitations of GenPCA-based trainings are highlighted and contrasted with instructor-led trainings.

### 5.1    Didactical methods for GenPCA-based training

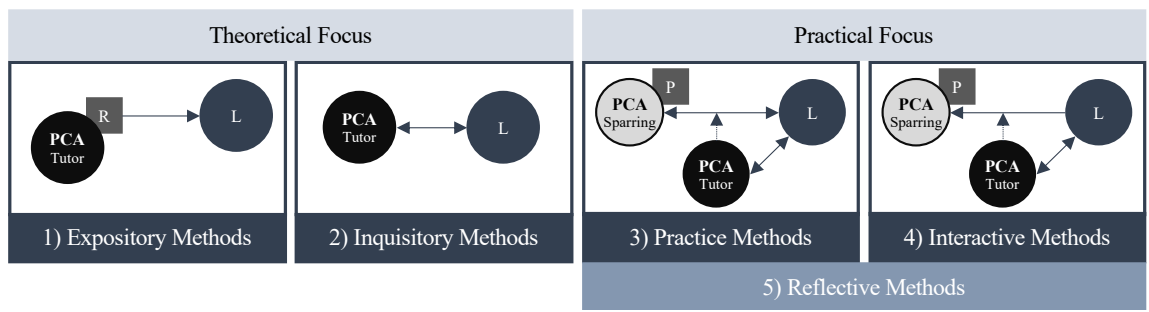

*Figure 2.*     Derived didactical methods for GenPCA-based training in workplace learning

When integrating a GenPCA into workplace learning to train the communication skills of employees, different didactical methods can be applied to support the learning and to design the learning process. These didactical methods should enhance various aspects of workplace learning, like self-directed learning, experiential learning, or adaptive learning, to ultimately improve the motivation of learners. As our results show, the didactical methods suitable for a GenPCA-based training can be split into theoretical and practical methods, as depicted in Figure 2. Depending on the specific context and objective of the workplace learning a mix of different methods can then be applied to integrate theoretical and practical aspects. As observed in the interviews with employees of two public administrations in Germany, who participated in a GenPCA-based training, five didactical methods are applicable and useful.

Concerning the presentation of knowledge, our GenPCA-based training supported *1) Expository Methods* and *2) Inquisitory Methods* when integrating DIMA as a tutor. With *1) Expository Methods*, learning resources (R), such as videos, texts, questions, or exercises, are exposed to the learners (e.g., via videos) without proactive presentation through or interaction with a GenPCA. However, such a GenPCA is aware of the learning resources and can still provide additional information or support to enrich the learning material. In contrast, when using *2) Inquisitory Methods*, a GenPCA directly interacts with the learners in the role of a tutor, e.g., to deepen the knowledge about a specific topic. In our study, the learners were able to watch training videos, which discussed different communicative aspects or were presented with learning questions. Even though DIMA did not present these materials directly to the learners, it was aware of the learning resources and provided additional information about them or better structured the presented material, e.g., by offering summaries about the videos. As observed in the interviews, the employees valued DIMA's abilities to individually reply to answers about the learning material and process. So, the natural answers were perceived as helpful and supportive. Previously, PCAs were commonly overwhelmed by questions and interactions that were unforeseen and not accounted for (Janssen et al., 2021; Khosrawi-Rad et al., 2023; Tolzin et al., 2023). This often leads to frustration and disturbs the learning process. However, DIMA's flexibility and ability to naturally explain learning material or to answer process-related questions enriched the learning experience. Hence, relying on GenAI to answer learners' questions allows for high-quality and human-like interactions similar to a human instructor. Using GenPCAs for expository and inquisitory methods is thus a feasible option to share knowledge with many learners in an engaging way and to unburden human instructors.

Additionally, GenPCA-based trainings can offer a set of practice-focused methods to train the workforce in communication skills. The didactical methods in this category include *3) Practice Methods* and *4) Interactive Methods*, which are best combined with *5) Reflective Methods*. By offering *3) Practice Methods* and *4) Interactive Methods*, GenPCA-based training can immerse learners in a learning situation that feels close to reality while providing a safe ground for learning and exercising. Although both methods involve the GenPCA in the role of a sparring partner, the focus on skill and knowledge activation differs. Whereas *3) Practice Methods* activate learners' skills and knowledge through conducting practical exercises (e.g., interacting with a simulated citizen), *4) Interactive Methods* require the learners to explain specific learning aspects (e.g., to a simulated colleague). Providing practical exercises is a fundamental aspect of workplace learning, as it connects skills and knowledge with practical experiences (Kolb and Kolb, 2009; Noe et al., 2010; Wang, 2018). However, organizing practical exercises is often challenging. They typically require careful preparation and human sparring partners, and often fail to fully immerse learners. In this study, DIMA engaged with employees as a sparring partner and offered practice and interactive exercises. As the learners reported, the realistic scenarios combined with DIMA's ability to adapt to conversational situations and express emotions, created an engaging and realistic training experience. The learners were able to immerse themselves in simulated learning situations that GenPCAs can establish with their dialog and speech-generation capabilities. By practicing their communication skills in such a setting, they felt more confident and confirmed in their skills. DIMA created a realistic testing ground for practicing communication skills and offered insights into various real-world scenarios, which were previously unknown to learners. Previous approaches to integrating PCAs into practice situations lacked the flexibility to respond to unpredictable situations and could not adjust their language to the context (Bucher et al., 2023). This often complicated their development and provided a less immersive experience. By leveraging LLMs, GenPCA-based trainings enable a wide range of scenarios via prompt engineering, which does not require expensive training of NLP models, and create a realistic learning experience like human-led training.

Evaluating learners' performances and providing supportive feedback to guide future improvement is a critical and challenging task for instructors (Ingram et al., 2013). It requires careful evaluation and also necessitates understandable and considerate communication of feedback. Based on *5) Reflective Methods*, the learners receive personalized and performance-based feedback from the GenPCA in the role of a tutor. In this study, DIMA provided feedback after the employees finished an exercise with the sparring partner. Therein, DIMA processed the entire dialog between the sparring partner and learner and evaluated it based on pre-defined criteria. This feedback was well perceived by the employees and created a better understanding of the learners' actual learning level and the learning material. DIMA's neutral character also created a relaxed learning atmosphere. In contrast to human-led training, learners did not fear being judged or evaluated. Similar effects were found in previous studies, which reported on PCA feedback or the power distance of the PCAs and its effects on learners (Heidig and Clarebout, 2011; Schlimbach and Zhu, 2023). By integrating reflective methods, GenAI-based trainings incorporate high-quality and adaptive feedback, which is not perceived as judgmental, but supportive.

## 5.2    Motivational aspects of GenPCA-based training

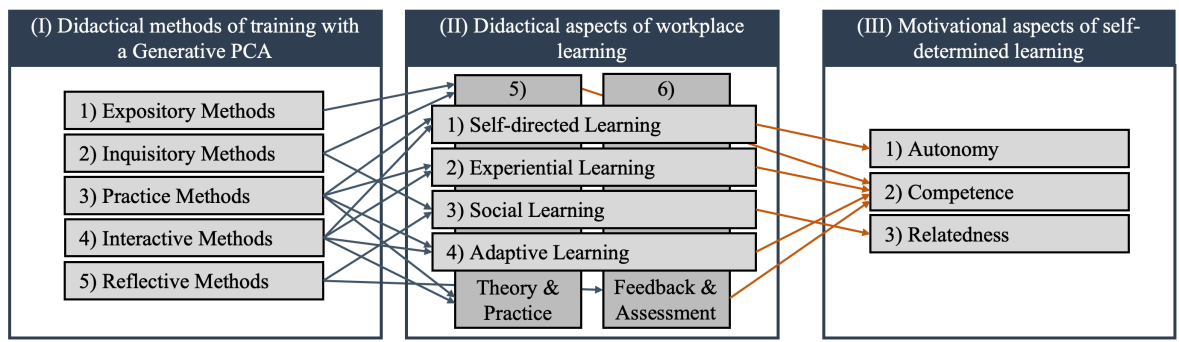

*Figure 3.        Mapping didactical methods and aspects with their effects on self-determination*

A fundamental challenge of workplace learning is to motivate employees. Establishing motivation requires careful consideration of the adult learners' needs for autonomy, competence, and relatedness (Ryan and Deci, 2000). These needs can be influenced by various didactical aspects of workplace learning, including self-directed learning or the integration of theory and practice. Whereas instructor-led training is well suited to personalize the learning material to the learners' competencies, they are less self-directed and rather inflexible in terms of time and location. Additionally, they often lack learner-centricity (Noe et al., 2010). In contrast, e-learning can be conducted from almost anywhere at any time but lacks adaptivity and practical exercises (Noe et al., 2010). As our findings show, a GenPCA-based training combines the benefits of both by offering self-determined learning for employees and establishing learner-centricity. As illustrated in Figure 3, integrating various didactical methods in a GenPCA-based training supports different didactical aspects of communication training for employees, like self-directiveness or adaptiveness. These foster the psychological needs of learners.

First, our findings indicate that a GenPCA-based training can positively influence the learners' *need for autonomy*. The need for autonomy of adult learners is mainly influenced by the self-directiveness of the training (Lemmetty and Collin, 2020; Taylor, 2006). *Self-directiveness* allows employees to take charge of their learning activities and the learning process. This is especially important for adult learning (Knowles et al., 2020). When ignoring this, learning with a PCA can quickly turn into perceived surveillance (Bucher et al., 2023). Conducting practical exercises with DIMA (practice and interactive methods) without the need for human involvement provided the employees with more freedom to select the time and location of their training activities and shifted the focus to the learner. Therein, GenPCAs leverage the capabilities of LLMs to provide authentic learning experiences, enhancing the capabilities of prior PCA generations in their functions as peers or sparring partners (Chen et al., 2020; Kim and Baylor, 2016; Tegos and Demetriadis, 2017). Hence, through the integration of GenPCAs, human intervention for conducting practical exercises or personalized feedback becomes obsolete in a GenPCA-based training, which promotes self-directedness and autonomy during the training.

Secondly, GenPCA-based training is well suited to support the adult learners' *need for competence*. Similar to human instructors, GenPCAs can provide adaptive and experiential learning, which helps to integrate theory and practice. Also, they can assess learners' performances during the training and offer individual feedback. These didactical aspects directly influence the learners' perception of competence (Karabenick, 2014; Kolb and Kolb, 2009; Noe et al., 2010; Plass and Pawar, 2020; Wang, 2018). For instance, by offering feedback based on the learners' performance during the exercises (reflective methods), DIMA was able to link the feedback with the learner's competence level instead of providing generic feedback. Further, by offering practical exercises, either practice or interactive-oriented, DIMA was able to establish an experiential learning environment. Practicing in such learning environment made the employees feel more confident in their skills and, thus, fulfilled their need for competence.

Lastly, GenPCA-based training influences employees' need for relatedness. Generally, embedding the learning activities in a social environment is crucial and determines peoples' perception of relatedness (Minnaert et al., 2011; Oh and Song, 2021). Using GenPCAs, a social learning environment can be established, and learners' need for relatedness met. People quickly project human values onto anthropomorphic systems and establish social bonds with them (Khosrawi-Rad et al., 2023; Lester et al., 1997). Integrating GenAI allows for enhanced natural dialog capabilities and social cues, which create a realistic and human-like perception of the GenPCA. This, in turn, evokes feelings of social presence and connection between learners and GenPCA. In this study, DIMA's abilities to converse in natural language over various modalities affected DIMA's social presence. Even though DIMA's voice was still sometimes perceived as robotic, it was still able to establish feelings of connection with many employees. Using GenPCAs to create a feeling of closeness and, hence, to meet peoples' need for relatedness, is useful when training a diverse and geographically dispersed workforce.

### 5.3     Limitations of GenPCA-based training for workplace learning

GenPCA-based training comes not without limitations. For instance, GenPCAs are not yet capable of detecting the learners' state of mind. In this study, the adaptation of exercises and interactions was primarily based on communicative aspects achieved by LLMs. However, this limits the possibilities of adapting the learning material and exercises to the true level of the learners' competencies and restricts the reaction, e.g., the adaptation of the learning process, to the learners' emotional states. When interacting with a human instructor, this instructor often implicitly senses and perceives subtle cues, e.g., through the learners' facial expressions, bodily movements, or language, that indicate comprehension, boredom overwhelming, or other cognitive and emotional states (Siu and Wong, 2016; Urhahne and Zhu, 2015). Human instructors can then react and adapt the learning process accordingly, e.g., by selecting tasks that fall into the zone of proximal development and provide the optimal setting for learning. Despite prior research on PCAs and their ability to cover the ZPD (Winkler et al., 2020), it remains open if current approaches to developing PCAs can fully grasp the learners' competencies and adapt the learning process appropriately. Similarly, human instructors commonly rely on their intuition to express empathy. For instance, when realizing that the learner is stressed or in a bad mood, the instructors can quickly react and adapt the learning situation. Even as research has begun investigating the use of image and video recognition to detect learners' state of mind (Mukhopadhyay et al., 2020), these approaches are still nascent and not integrated into current LLMs. It also remains questionable if the detection of subtle cues that require human intuition can be achieved at all and if employees accept such analyses.

Our results also unveiled that GenPCAs lack authority and responsibility over the learning process. For human instructors, it is a balancing act to provide as much self-directiveness and autonomy as possible while still conveying the importance of the training and keeping learners active (Noe et al., 2010). If needed, human trainers can actively deploy their authority to manage the learning process. However, DIMA lacked such authority and could not intervene and enforce the training. We believe that this issue cannot solely be solved by the technical design of a GenPCA. Instead, organizations need to consider how GenPCAs can be integrated into the organizational structure. For instance, GenPCAs could augment instructor-led training with their didactical methods. As a result, they would widen the reach of instructor-led training to a large workforce while combining various benefits of GenPCA-based and instructor-led training. In the educational sector, there exists research that focuses on augmenting teachers with PCAs (Sjöström and Dahlin, 2020). It would be interesting to extend their scope to the organizational context and the didactical methods of GenPCA-based training.

## 6     Conclusion

GenPCA-based training, building on LLMs, can be used to systematically train employees in a workplace setting. The evaluation of our prototype DIMA showed promising results: the agent was well received by employees and contributed to self-determined learning. The agent established learner-centricity by providing a self-directed, adaptive, experiential, and social learning environment that integrated theory and practice and offered individualized feedback. This was achieved by the capabilities of our GenPCA, which guided the learner through the learning process by responding to learners' queries, by engaging with learners in practical exercises, and by offering personalized and corrective feedback. The results illustrate that through the integration of five didactical methods, GenPCAs can successfully implement common approaches of workplace learning and enable self-determined learning. Despite smaller issues concerning the quality of voice generation and limitations of GenPCAs to understand the learners' state of mind, e.g., to fully leverage the zone of proximal development, GenPCAs can fill the gap of human trainers. They can make an engaging learning environment, similar to instructor-led training, accessible for a large number of employees without the costs and efforts induced by human trainers.